\documentstyle[11pt,aasms]{article}
\lefthead{??}
\righthead{}

\def\lta{{\>\rlap{\raise2pt\hbox{$<$}}\lower3pt\hbox{$\sim$}\>}}
\def\gta{{\>\rlap{\raise2pt\hbox{$>$}}\lower3pt\hbox{$\sim$}\>}}

\lefthead{??}
\righthead{}
 
\begin{document}

\title {HST Optical-NIR Colors of Nearby $R^{1/4}$ and Exponential Bulges\footnote{Based on observations with the
NASA/ESA Hubble Space Telescope, obtained at the Space Telescope
Science Institute, which is operated by Association of Universities
for Research in Astronomy, Inc.\ (AURA), under NASA contract
NAS5-26555}}

\author{C. Marcella Carollo\footnote{Visiting Astronomer, The Johns
Hopkins University, Department of Physics and Astronomy, Baltimore MD
21218}} \affil{Columbia University, Department of Astronomy, 538
W.\ 120$^{th}$ St., New York, NY 10027}

\author{Massimo Stiavelli} \affil{Space Telescope Science Institute,
3700 San Martin Drive, Baltimore MD 21218}

\author{P. Tim de Zeeuw} \affil{Sterrewacht Leiden, Postbus 9513, 2300
RA Leiden, The Netherlands}

\author{Marc Seigar \footnote{Visiting Astronomer, Space Telescope
Science Institute, Baltimore MD 21218} \affil{Sterrenkundig
Observatorium, Universiteit Gent, Krijgslaan 281, B-9000 Gent,
Belgium}}

\author{Herwig Dejonghe} \affil{Sterrenkundig Observatorium,
Universiteit Gent, Krijgslaan 281, B-9000 Gent, Belgium}

\begin{abstract}

\noindent
We have analysed $V$, $H$ and $J$ HST images for a sample of early- to
late-type spiral galaxies, and reported elsewhere the statistical
frequency of $R^{1/4}$-law and exponential bulges in our sample as a
function of Hubble type, and the frequency of occurrence and
structural properties of the resolved central nuclei hosted by
intermediate- to late-type bulges and disks (see references in the
text).  Here we use these data to show that: {\it (i)} The $V-H$ color
distribution of the $R^{1/4}$ bulges peaks around $<V-H> \sim 1.3$,
with a sigma $\Delta (V-H) \sim 0.1$ magnitudes. Assuming a solar
metallicity, these values correspond to stellar ages of $\approx 6 \pm
3$ Gyrs.  In contrast, the $V-H$ color distribution of the exponential
bulges peaks at $<V-H> \sim 0.9$ and has a sigma $\Delta (V-H) \sim
0.4$ mags. This likely implies significantly smaller ages and/or lower
metallicities for (a significant fraction of the stars in) the
exponential bulges compared to the $R^{1/4}$-law spheroids. {\it (ii)}
Most of the central nuclei hosted by the exponential bulges have $V-H$
and $J-H$ colors which are compatible with relatively unobscured
stellar populations. Assuming no or little dust effects, ages $\gta 1$
Gyrs are suggested for these nuclei, which in turn imply masses of
about a few 10$^6$ to a few 10$^7$ M$_\odot$, i.e., sufficient to
dissolve progenitor bars with masses consistent with those inferred
for the exponential bulges by their luminosities. {\it (iii)} While
different bulge-nucleus pairs cover a large range of $V-H$ colors,
each bulge-nucleus pair has quite similar $V-H$ colors, and thus
possibly similar stellar populations. The HST photometric analysis
suggests that exponential-type bulge formation is taking place in the
local universe, and that this process is consistent with being the
outcome of secular evolution processes within the disks.  The
structures which are currently formed inside the disks are quite
dissimilar from the old elliptical-like spheroids which are hosted by
the early-type disks.
\end{abstract}

{\it subject headings}: galaxies: formation - galaxies: evolution -
galaxies: structure - galaxies: nuclei - galaxies: bulges

\section{Introduction}

\noindent
The size of the central bulge compared to the size of the disk is one
of the classification criteria of the entire Hubble sequence (Sandage
\& Bedke 1994). Furthermore, bulges host a large fraction of the
baryons which have been converted into stars during the Hubble time
(e.g., Fukugita, Hogan \& Peebles 1998). It is also likely that the
formation of bulges plays a role in polluting the intergalactic medium
with processed matter and radiation, and thus affects the formation of
luminous structure on larger scales.  Yet, to date, it is not
understood not only `how' bulges form, but even whether they form
before, contemporaneously with, or after the disks (e.g., see the detailed
review by Wyse, Gilmore \& Franx 1997, and also Bouwens, Cayon \& Silk 1999). 
In the one extreme, bulges may arise from the collapse of a primordial gas
cloud into clumps which then merge together, and the disk be a later
accretion by gas infall (e.g., Eggen et al.\ 1962; Larson 1975;
Carlberg 1984). In the opposite extreme, bulges may form by vertical
dynamical instabilities of the disks (e.g., Combes et al.\ 1990;
Pfenniger 1993).

Historically, `bulges' have been recognized as such because they
appear as high surface brightness, concentrated objects in the
considerably larger and fainter disks.  These `bulges by definition'
often also have quite a distinct color, and, as the name suggests, are
considerably fatter than the disks.  Superficially, they appear to
share many properties with small elliptical galaxies: The relatively
few and typically rather massive disk-embedded spheroids which have
been studied in detail show an $R^{1/4}$ radial light profile,
relatively old and metal-enriched stellar populations, and are
rotationally-supported systems (Wyse et al.\ 1997 and references
therein).  Despite the temptation of historical inertia, clearly
wanting us to restrict the use of the name `bulge' to this
sort-of-well-defined class of stellar systems only, it is however
unwise, in the quest for clarifying the origin of these fundamental
components of one-Hubble-time-old galaxies, to ignore that different
kinds of stellar structures are also often found in the centers of the
local-universe disks (Wyse et al.\ 1997).  

Significant differences between the `bulges' of intermediate- to
late-type disks and the more massive spheroids have in fact been
found.  First, preliminary studies suggest that the stellar
populations of small bulges in general may be younger and more
metal-poor than those of the large bulges (Trager et al.\ 1999).
Furthermore, several studies have revealed `pseudo-bulges' with e.g.,
cold kinematics (Kormendy 1993), a peanut-shape morphology associated
with bar-like kinematics (e.g., Kuijken \& Merrifield 1995; Bureau \&
Freeman 1999) or an exponential -- rather than $R^{1/4}$ -- fall-off
of the light distribution (hereafter exponential bulges; e.g.,
Kormendy \& Bruzual 1978; Shaw \& Gilmore 1989;
Courteau et al.\ 1996 and references therein).

The pseudo-bulges also appear as clearly distinct structures from the
disks, and replace the `classical' bulges in disks that, at least
superficially, look rather similar to others that host instead the
canonical $R^{1/4}$ structures. It is unclear how these strange
central structures relate to their $R^{1/4}$ relatives. However, the
fact that they `substitute for' the $R^{1/4}$ bulges in otherwise
normal intermediate-type disks provides an argument for including them
in any thorough exploration which is aimed at clarifying the origin
and evolution of the spheroidal components of the local disk
galaxies. These pseudo-bulges may lead us to understand by comparison
which are the required circumstances for forming a dense
elliptical-like spheroid (failing which, only a pseudo-bulge is
formed), or may indicate the occurrence of fundamental changes with
cosmic time in the evolutionary paths of galaxies, or may turn out to
be consistent with being evolutionary related to the `classical'
bulges for which we seek the origin and progenitors.

In order to help clarify the origin of the structural diversity
between the largest and the smallest disk-embedded spheroids,
we performed a Hubble Space Telescope (HST) snapshot survey of $\approx
80$ spiral galaxies randomly-selected out of a complete sample of 134
targets. The central regions of the selected objects were imaged in
the visual (with {\tt WFPC2} and the F606W filter) and in the
near-infrared (NIR; with {\tt NICMOS} and mostly the F160W and
occasionally the F110W filter). The observations and the data
reduction are fully described in a series of papers,
separately for the {\tt WFPC2} survey (Carollo et al.\, 1997; Carollo
et al.\, 1998; Carollo \& Stiavelli 1998; Carollo 1999) and the {\tt
NICMOS} survey (Carollo et al.\, 2000, paper I; Seigar et al.\, 2000,
paper II). The high angular resolution of the HST data allowed us to
mask out from several of the images patches of dust and knots of star
formation, and thus derive for those targets `clean' isophotal models
for the underlying galaxian light.  Two-component `bulge plus disk'
analytical fits were applied to the so-derived radial light profiles.
Echoing previous voices, we considered as a `bulge' that distinct
central structure which `contains all the light in excess of the
inward extrapolation of a constant scalelength exponential disk' (Wyse
et al.\ 1997; Gilmore 1999). Within this more general definition of
bulge, we found that the bulge light profile was typically best
modeled with an $R^{1/4}$-law in the early-type disks and with an
exponential law in the later types, in agreement with previous
studies (e.g., Courteau et al.\ 1996).

In this paper we combine the NIR measurements presented in papers
I and II with the {\tt WFPC2} measurements so as to investigate the
optical-NIR colors of the exponential versus $R^{1/4}$-law bulges
found in our HST survey, as well as the colors of the
photometrically-distinct nuclei which we found embedded in the
dynamical centers of all the exponential bulges. The objects included
in this study are listed in Table 1, together with the information on
whether they have an $R^{1/4}$-law or an exponential bulge, on the
$V-H$ color for the host bulges, and on the $V-H$ color for the
embedded nuclei.  The intrinsic dust-age-metallicity degeneracy of
broad-band colors for intermediate-to-old age stellar populations is a
well-known problem. Nonetheless, although the calibrations of absolute
ages and metallicities do remain elusive, photometric indicators can
still be a powerful bench-mark for `ranking' the stellar populations
among diverse kinds of bulges, and among bulges and other galactic
sub-components such as the photometrically-distinct nuclei.  We
briefly discuss plausible implications of our results for the
formation of bulges during the course of cosmic history.

\section{Results}

\subsection{The Colors of Exponential and $R^{1/4}$-law Bulges}

\noindent
The majority of the $R^{1/4}$-law bulges found in our sample are
bright (massive) systems embedded in early-type disks, in contrast
with the fainter exponential bulges which are found mostly in Sb to Sc
hosts, in agreement with previous studies (see Figure 1, where the absolute
$V$ magnitude of exponential and $R^{1/4}$-law bulges is plotted
versus their $V-H$ colors).  Figure 2 shows the $V-H$ color
distribution for the $R^{1/4}$-law (solid line) and exponential
(dashed line) bulges detected in our HST survey. The measurements are
expressed in $AB$ magnitudes. The colors are obtained by integrating
the smooth analytical fits to the bulge light profiles inside $1/2
R_e$ (or within $6''$ if that radius exceeds the radial extension of
the {\tt NICMOS} data); therefore they exclude (or at least minimize)
any contribution from e.g., the nuclei, strong dust lanes/patches and
knots of recent star formation.

The $V-H$ color distribution of the $R^{1/4}$ bulges peaks around
$<V-H> \sim 1.3$, with a sigma $\Delta (V-H) \sim 0.1$
magnitudes. Bruzual \& Charlot models, calibrated into the F606W and
F160W filters with Synphot/IRAF, indicate that the average metallicity
of these systems must be at least half-solar for their age to be
smaller than that of the universe. Assuming a solar metallicity, the
average $<V-H>$ color corresponds to an average age for the population
of $\approx6$ Gyrs, with a spread of about 3 Gyrs around this
average. The presence of a small fraction of younger stars cannot be
clearly excluded on the basis of the available data alone.  In the
absence of dust, the bluest colors that we observe in the R$^{1/4}$
bulges could be obtained by contaminating, e.g., a 9 Gyrs old
underlying population with a 10\%, 1\% or 0.1\% in mass of stars 1
Gyr, 100 Myrs or less-than-5 Myrs old, respectively.  Possible
`solutions' clearly exist also in the presence of dust: Using the
`screen-approximation' and a Cardelli extinction law (Cardelli,
Clayton \& Mathis 1989) as a benchmark, with e.g., an extinction in
the $V$ band of $A_V = 3$, about 10\% (in mass) of the population
should be less than 5 Myrs old in order to reproduce the bluest end of
the $R^{1/4}$ bulges $(V-H)$ distribution.  A more modest $A_V=1.5$
could accomodate a fraction as high as $\sim 70\%$ of young stars, if these
were about 100-Myrs-old.  On the other hand, the smooth morphology of
this class of objects and their color maps, virtually featureless at
all radii, make it rather implausible that these systems contain large amounts of dust.  In fact, the only features detected in the
color maps for the $R^{1/4}$ bulges are some mostly-nuclear patches of
dust. The effects of these dust patches are removed by measuring the
colors by integrating over the light profiles (since the latter are
obtained by means of isophotal fits in which the patches of dust are
masked out).  The most plausible solution for the $R^{1/4}$ bulges is
therefore that at least more than 90\% of their stars are older than
several Gyrs.  This finding echos previous results for the
$R^{1/4}$-law bulges (e.g., Peletier et al.\ 1999 and references
therein) by supporting the idea that many of these dense spheroids
are, at least population-wise, similar to small ellipticals.

The $V-H$ color distribution of the exponential bulges, in contrast,
presents some surprises. The average $<V-H>$ of these systems is in
fact $\sim 0.4$ magnitudes bluer than the corresponding average for,
and the sigma $\Delta (V-H) \sim 0.4$ magnitudes is significantly
larger than that of, the $R^{1/4}$ bulges. Furthermore, although a few
of the exponential bulges are actually as red as the reddest of the
$R^{1/4}$-law bulges, the former extend to significantly bluer colors
than the latter, down to a $(V-H) \sim 0$. In principle, adding a
small fraction of young stars to an underlying old stellar population
could again explain the observed colors. For example, in the absence
of dust, a 10\% of 100 Myr old stars mixed to the remaining 9-Gyr-old
population would be sufficient to obtain a $(V-H) \sim 0.5$
magnitudes; the required percentage of young stars would clearly
decrease for even younger stellar populations. In the presence of
substantial amounts of dust, say $A_V$ up to 2 magnitudes, the age of
the population should decrease down to $\lta 5$ Myr in order to obtain
a $(V-H) \sim 0.5$ with a 10\% or smaller contamination of young
stars.  Still, thanks to the high angular resolution of the HST images
which allow to perform an accurate masking of any localized feature in
the images, the effects of dust patches and possible knots of star
formation are also in this case negligible in the colors derived from
the light profiles (from isophotal fits). Therefore, while a diffuse
younger component well-mixed with the underlying old population cannot
clearly be ruled out by the $(V-H)$ color alone, the most appealing
solution for the exponential bulges is that they are on average
significantly younger than their $R^{1/4}$ relatives.

\subsection{Colors of the Nuclei}

\noindent
About a third of our sample was imaged in all three filters $J$, $H$
and $V$. In Figure 3 we plot the $J-H$ color versus the $V-H$ color
for the compact nuclei found embedded in intermediate- to late-type
galaxies, including the nuclei which are found in the centers of the
exponential bulges. Actually, each exponential bulge in our sample
hosts one such a nucleus (Carollo 1999).  In the Figure, squares are
the nuclei embedded in the exponential bulges. The triangles represent
instead the nuclei embedded in systems with no isophotal fit, i.e.,
systems whose central regions are heavily obscured by dust or
complicated by recent star formation, so that for them no reliable
isophotal fit could be performed (see previous papers for the
corresponding list of measurements).  Dust extinction affects points
on this diagram by shifting them along the arrow shown in the
upper-left corner of the Figure.  The arrow is drawn for a 1 mag
extinction in the $V$ band.  Theoretical tracks derived from Bruzual
\& Charlot models for metallicities $Z=0.02Z_\odot$ (dotted line) and
$Z=Z_\odot$ (solid line) are also shown in Figure 3.  Stellar ages
increase upward along the tracks from $\approx$100 Myr to 18 Gyr.

Many central compact nuclei (typically in hosts without an analytical
fit to the bulge) have $V-H$ and $J-H$ colors which are incompatible with
arising from unobscured stellar populations. These nuclei are 
embedded in very complex circum-nuclear structures, e.g., in strong
dust lanes/arms or rings/arms of recent star formation. For a few of
these systems information on the central spectral properties is
available at ground-based resolution, i.e., at a resolution covering
an area $\approx100$ times larger than that covered by the nuclei, an
area which includes this complex circum-nuclear structure as well
(making the spectra not suitable for studying the physical properties
of the central compact nuclei, for which we have to rely on the HST
photometry only). On ground-based scales of $\approx$1kpc, these
galaxies show an active-type central spectrum, either of AGN- or of
HII-type.  It is therefore not surprising that the optical/NIR light
from their central resolved nuclei can also be polluted by several
magnitudes of extinction, and/or can contain non-thermal emission from a
central AGN.

In contrast, most of the central nuclei which are embedded in the
exponential bulges have $V-H$ and $J-H$ colors which are compatible
with arising from nearly unobscured stellar populations.  Spectral
information is not available for most of their host galaxies, not even
from the ground. In principle, large amounts of dust extinction
cannot be ruled out even for these nuclei. However, in contrast with
the complexity of structure generally underlying the central nuclei
with very red $V-H$ and $J-H$ colors, the exponential bulges are
generally rather smooth, amorphous systems, showing not nuch evidence
for the presence of large amounts of dust. This supports the idea that
the optical/NIR light of the nuclei of exponential bulges is mostly
stellar, i.e., it is not heavily polluted by a non-thermal component
or by large amounts of dust. Under this assumption, relatively old
ages -- of the order of $\approx$ 1Gy and above -- are suggested by
the combined $V-H$ and $J-H$ colors for several of these nuclei
embedded in the exponential bulges.

If spectroscopically confirmed for at least some of these nuclei,
these relatively old ages would have interesting consequences for the
masses of these systems.  In Figure 4 we plot the absolute magnitude
$V$ against the $V-H$ color for all the nuclei for which both a $V$
and an $H$ image are available (for several of these a $J$ image is
not available; this is the reason why there are more points here
compared to Figure 3).  Age increases from top to bottom along the
plotted Bruzual \& Charlot tracks; these refer to a $2.5 \times 10^6
M_\odot$ star cluster of metallicities $Z=0.02Z_\odot$ (dotted line)
and $Z=Z_\odot$ (solid line). Dust extinction shifts points along the
direction indicated by the upper-left arrow.  The mass of a stellar
cluster and the age of its stellar population have nearly fully
degenerate effects on a color-magnitude diagram: The tracks can be
shifted upward (downward) by increasing (decreasing) the mass of the
star cluster (a 2.5 magnitudes shift implies a factor 10 variation in
mass); correspondingly, a larger (lower) stellar age will be
associated with any fixed point encompassed by the tracks on the $V$
vs $V-H$ diagram.  In this enlarged sample, again most of the central
compact nuclei which are embedded in the complex circum-nuclear
structure of those hosts without an analytical fit have $V-H$ colors
which are incompatible with arising from unobscured/unpolluted stellar
populations.  And again most of the nuclei embedded in the exponential
bulges are instead consistent with being relatively unobscured stellar
structures.  They are also, at face value, more enriched than e.g.,
the Milky Way globular clusters, since they typically lie rightward of
the metal-poor stellar track. If the large ages estimated from the
$(J-H)$--$(V-H)$ diagram are assumed for (some of) these nuclei, their
location on the $V$--$(V-H)$ plane implies masses in the thereabouts
of a few 10$^6$ to a few 10$^7$ M$_\odot$, depending on the exact
dating of the stellar structure between $\approx 1$ and many Gyrs.

\subsection{Comparison between Nuclei and Host Exponential Bulges}

\noindent
In Figure 5 we show the comparison between the $V-H$ color of the
exponential bulges and the $V-H$ color of their hosted nuclei. Within
the error bars there is a positive trend between the colors of the
nuclei and those of their host bulges (Spearman's and Kendall's rank
order tests give a probability of correlation greater than 94\% and
95\%, respectively). The slope of the correlation is compatible with
being unity, but the error bars are too large to determine it
reliably. Different bulges-nuclei pairs cover a large range of $V-H$
colors (consistently with what discussed in Figure 2), but most
bulge-nucleus pairs seem to have  possibly rather similar $V-H$
colors. Conspiracies may be at play; on the other hand, this result
suggests that the nuclei embedded in the exponential bulges may have
stellar populations similar to those of their host bulges.

\section{Discussion}

\noindent 
It is by now a decade ago that  e.g., Hasan \& Norman (1990), Norman \&
Hasan (1990) and Pfenniger \& Norman (1990), proposed some basic
dynamical mechanisms that produce an amplification of the accretion
rate of gas clouds into the central regions of barred galaxies, and
discussed their effects on the evolution of disk galaxies and their
implications for bulge formation.  Only a few years later John
Kormendy presented the first evidence for cold kinematics in
intermediate-type bulges at the IAU Symposium 153, thereby providing
extremely compelling observational ``evidence that some bulges are really
disks'' (Kormendy 1993). Since then, several other studies have been
performed that further support a close connection between bulges and
disks in Sb-Sc galaxies (e.g., Kuijken \& Merrifield 1995; Bureau \&
Freeman 1999; Courteau et al.\ 1996 and references therein), and thus
possibly the growth of bulges inside the disks due to the secular
evolution of the latter.

The theoretical suggestions for how the `disk-driven' bulge formation
may take place invoke the presence of non-axisymmetric perturbations,
i.e. `bars', generated by some kind of disk instabilities. First, the
stars initially in a bar can be kicked off the plane of the disk by
 `buckling' instabilities, and form a three-dimensional bulge
(e.g., Combes et al.\ 1990). Furthermore, numerical experiments 
indicate that a very efficient way to induce significant evolution
along the Hubble sequence is to accrete mass in the centers of
galaxies hosting a tumbling triaxial structure such as a bar. The
addition of a central mass concentration of $\approx 1\%$ of the total
mass in fact disrupts orbits essential to its existence (e.g.,
Pfenniger \& Norman 1990; Hasan, Pfenniger \& Norman 1993; Norman,
Sellwood \& Hasan 1996). Bars do greatly enhance the efficiency of
dissipative mass transfer to the centers of the disks (Shlossman 1994;
Shlosman \& Robinson 1995; Shlosman, Begelman \& Frank 1990).  Thus,
provided there is an efficient mechanism for funneling the dissipative disk
material down to the very center, the bar itself is dissolved, and its
stars are deflected out of the plane of the disk, giving birth to a
three-dimensional bulge-like structure.

The broad-band photometric analysis presented here is clearly
afflicted by the well-known full-degeneracy of mass, age,
metallicity, dust, emission lines. The arguments of plausibility which
lead to the age and mass estimates for the exponential bulges and
their central nuclei discussed above are indeed to be taken with care:
Spectroscopic confirmation is needed. Still, within the 
caveats, those assumptions are a plausible solution with several
implications for the formation of bulges as a function of
cosmic time.

First, if dust effects are not at play, the fact that some exponential
bulges appear to be as red as the elliptical-like bulges suggests that
this `mode' of bulge formation may have been active even at the early
stages in our universe. However, the fact that the exponential bulges
appear to be bluer on average than the elliptical-like bulges does
suggest that, as a class, they are younger (and likely less
metal-enriched) than the latter. Therefore, on the one hand these new
observations support the idea that the growth of central structure
within the disks is still going on in our universe. On the other hand,
this growth seems to occur in the form of `exponential-type' structure
only, i.e., a structure quite dissimilar from the dense $R^{1/4}$-law
relics of the early times.  

The mechanisms that grow the central exponential bulgelike structures
remain clearly indetermined. However, at face value, the
`photometrically-estimated' central masses of the central nuclei
embedded in the exponential bulges are well-matched to those which
would have been required, in the bar formation/dissolution scenario,
to disrupt progenitors bars with masses of about a few $10^7$-$10^9
M_\odot$. These masses are consistent with the masses inferred for the
exponential bulges from their total $V$ luminosities (assuming similar
$M/L_V$ ratios for the exponential bulges and their nuclei; see also
Carollo 1999). 

The above arguments clearly do not prove that the bar
formation/dissolution mechanism operates: For example, the central
nuclei may have grown inside the exponential bulges due to dynamical
friction of globular clusters, as suggested by Tremaine, Ostriker \&
Spitzer (1975) for the nucleus of M31. At a time $t$, any globular
cluster within a radius $r \propto (Gmt/\sigma)^{1/2}$ will fall into
the center dragged by dynamical friction. For plausible values of $m$,
the mass of the infalling globular cluster, and $\sigma$, the velocity
dispersion of the exponential bulge,\footnote{No kinematic data are yet
available for the exponential bulges, so that `plausible' in this
context means `as derived from the Faber-Jackson (1976) relation and
allowing for generous ($\approx 50\%$) variations around the
Faber-Jackson value.} all the globular clusters within a few to
several half-light radii of the exponential bulges should have
converged to their centers in about 1 to a few Gyrs. The range of
masses inferred from the broad-band colors imply that about 10--100
globular clusters should have formed within such a radial range, a
value which is not ruled out given that many more globular clusters
will have formed than those which would be currently observed (and
given that the low central stellar densities inferred for the
exponential bulges, would allow the newly-borned clusters to sink in
the galaxy centers without being tidally-destroyed during their
infall, see Carollo \& Stiavelli 1998). 

On the other hand, the fact that the nuclei appear to be typically
more enriched than the Milky Way globular clusters, and the similar
$V-H$ colors of the nuclei and their host exponential bulges, does
argue in favour of similar stellar populations and thus of a
synchronous formation of these two galactic sub-components. A natural
way to synchronously form the exponential bulge and its central
nucleus would indeed be a scenario where the stars in the bulge and
the central nucleus are grown together by dissipative material
infalling into the galaxy center.

In summary, the high-resolution photometric analysis of
intermediate-type bulges, while far from being `conclusive', does
indicate that `non-classical' bulge formation is taking place in the
universe around us, and that this bulge formation is consistent with
arising from secular evolution processes within the disks. In
contrast, there is growing evidence that the early-type, massive
bulges not only are `old', but may even be as old as the Coma cluster
ellipticals, with an internal age-spread of only $\approx 2$ Gyr
(Peletier et al.\ 1999). Together, these results suggest that the
processes that have formed the dense elliptical-like spheroids in the
centers of the early-type disks are quite dissimilar from the
processes which are growing the disk-like central structures within
the later-type disks today.

\bigskip

\acknowledgements This research has been partially funded
 by Grants GO-06359.01-95A and GO-0731.02-96A awarded by STScI, and has made use of the
 NASA/IPAC Extragalactic Database (NED) which is operated by the Jet
 Propulsion Laboratory, Caltech, under contract with NASA.

\newpage

\begin{figure}
\caption{Absolute magnitude versus $V-H$ color for exponential (filled
squares) and $R^{1/4}$ (empty circles) bulges (AB magnitudes). 
The typical error bar is given for reference.}
\end{figure}

\begin{figure}
\caption{The $V-H$ distribution for exponential (dashed line) and
$R^{1/4}$ (solid line) bulges (AB magnitudes).}
\end{figure}

\begin{figure}
\caption{$J-H$ vs $V-H$ color-color diagram for the compact nuclei (AB
magnitudes). Squares are the nuclei embedded in the exponential bulges
(see Table 1). The triangles are the nuclei embedded in systems with
no isophotal fit (measurements reported in Carollo et al.\ 2000).
Tracks refer to Bruzual \& Charlot models of metallicities equal to
0.02$Z_\odot$ (dotted line) and $Z_\odot$ (solid line). The upper-left
arrow shows the effects of one magnitude of dust extinction in
$V$. The typical error bar is plotted in the  upper-left corner.}
\end{figure}

\begin{figure}
\caption{$V$ vs $V-H$ color-color diagram for the compact nuclei (AB
magnitudes). Symbols are as in Figure 3.  The Tracks refer to Bruzual
\& Charlot models for a $2.5 \times 10^6 M_\odot$ star cluster of
metallicities 0.02$Z_\odot$ and $Z_\odot$. The typical error bar is
plotted in the bottom-right  corner.}
\end{figure}

\begin{figure}
\caption{$V-H$ color of the exponential bulges versus $V-H$ color of
their own compact nuclei(AB magnitudes). The typical error bar is
also plotted. }
\end{figure}

\newpage

\begin{table*}
{\tiny\begin{center}\begin{tabular}{lcccc|lcc}
\hline\hline
\multicolumn{1}{l}{Name (Expo)} &
\multicolumn{1}{c}{$M_V$ (Bulge)}&
\multicolumn{1}{c}{$(V-H)_{Bulge}$} &
\multicolumn{1}{c}{$(V-H)_{Nucleus}$} &
\multicolumn{1}{c}{$(J-H)_{Nucleus}$} &
\multicolumn{1}{l}{Name ($R^{1/4}$)}& 
\multicolumn{1}{c}{$M_V$ (Bulge)}&
\multicolumn{1}{c}{ $(V-H)_{Bulge}$ }\\
\hline \tiny
ESO240G12	&     -18.6  &     -	&	1.0	$\pm0.6$ &	0.3     $\pm0.4$ &  NGC488      &     -22.6   & 1.38$\pm$0.09 \\
ESO498G5      	&     -17.3  & 1.21$\pm$0.12 &	1.3	$\pm0.3$ &	-	         &  NGC2196	&     -21.7   & 1.15$\pm$0.04 \\
ESO499G37      	&     -17.5  & 0.24$\pm$0.15 &	0.4     $\pm0.8$ &	-		 &  NGC2344     &     -21.7   & 1.13$\pm$0.09 \\
ESO548G29      	&     -18.4  &     -	&	1.2	$\pm0.7$ &	-	         &  NGC2460     &     -18.5   & 1.42$\pm$0.14 \\ 
ESO549G18     	&     -15.1  &     -	&	3.0	$\pm0.7$ &	0.4	$\pm0.4$ &  NGC3898     &     -20.7   & 1.36$\pm$0.12 \\ 
ESO572G22       &     -17.7  &     -	&	0.8	$\pm0.4$ &	0.2	$\pm0.3$ &  NGC3900     &     -18.7   & 1.19$\pm$0.04 \\ 
NGC406    	&     -19.0  & 0.81$\pm$0.08 &	0.7	$\pm0.4$ &	0.2	$\pm0.3$ &  NGC5985     &     -19.4   & 1.28$\pm$0.06 \\ 
NGC1345  	&     -18.2  & 0.41$\pm$0.09 &	0.8	$\pm0.3$ &	-	         &  NGC6340     &     -19.1   & 1.23$\pm$0.15 \\ 
NGC1483   	&     -17.8  & 1.10$\pm$0.09 &	0.2	$\pm0.5$ &	-	  &         NGC7280     &     -19.4   & 1.17$\pm$0.05 \\ 
NGC2082		&     -16.0  & 0.92$\pm$0.08 &	1.7	$\pm0.7$ &	0.5	$\pm0.4$ &&&\\
NGC2758		&     -17.2  & 0.41$\pm$0.07 &	0.2	$\pm0.7$ &	-	&&&\\
NGC3259		&     -15.5  & 1.14$\pm$0.13 &	1.1	$\pm0.7$ &	-	&&&\\
NGC3455		&     -13.8  & 0.95$\pm$0.18 &	1.3	$\pm0.4$ &	0.4	$\pm0.3$ &&&\\
NGC4980		&     -18.8  & 0.66$\pm$0.14 &	0.6	$\pm0.4$ &	0.2	$\pm0.3$ &&&\\
NGC6384		&     -17.6  & 1.56$\pm$0.09 &	2.1	$\pm0.9$ &	0.3	$\pm0.4$ &&&\\
\hline
\end{tabular}\end{center}}
\caption{The sample of exponential (left side) and $R^{1/4}$-law
(right side) bulges included in this study. The measurements, in $AB$
magnitudes, are from Carollo et al.\ 1997, Carollo et al.\ 1998 and
Carollo et al.\ 2000.}
\label{tab1}
\end{table*}
\normalsize

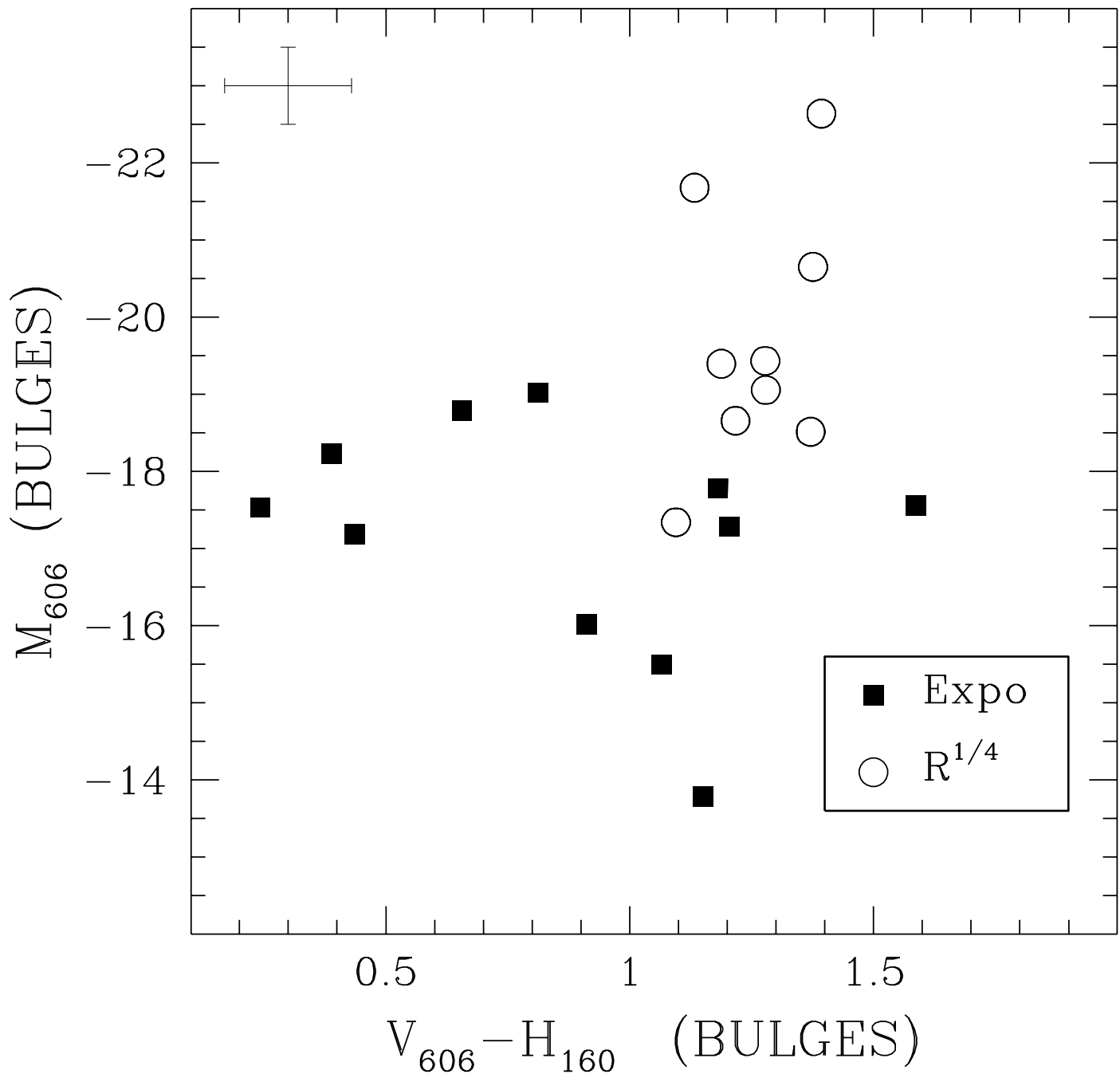
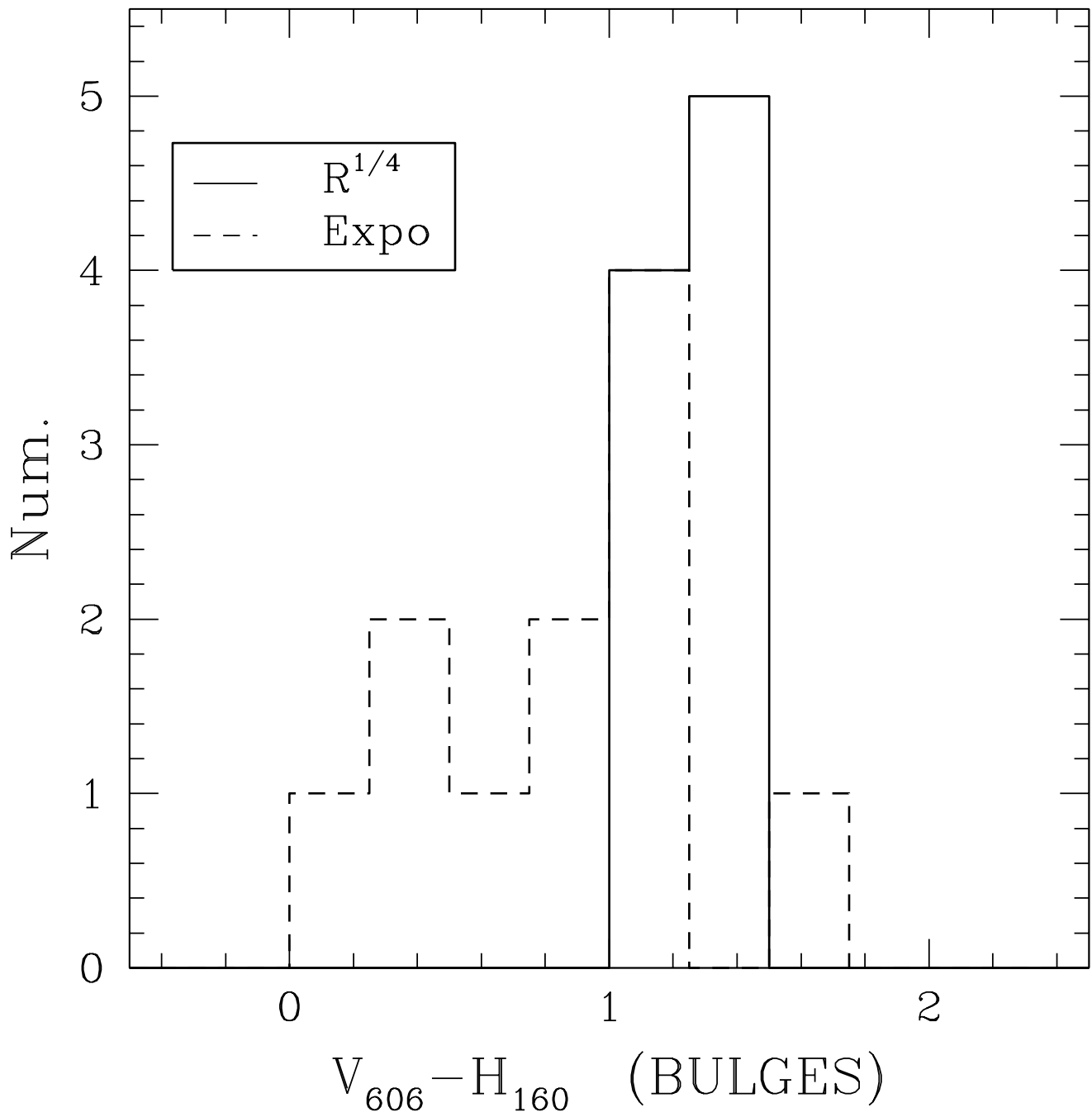
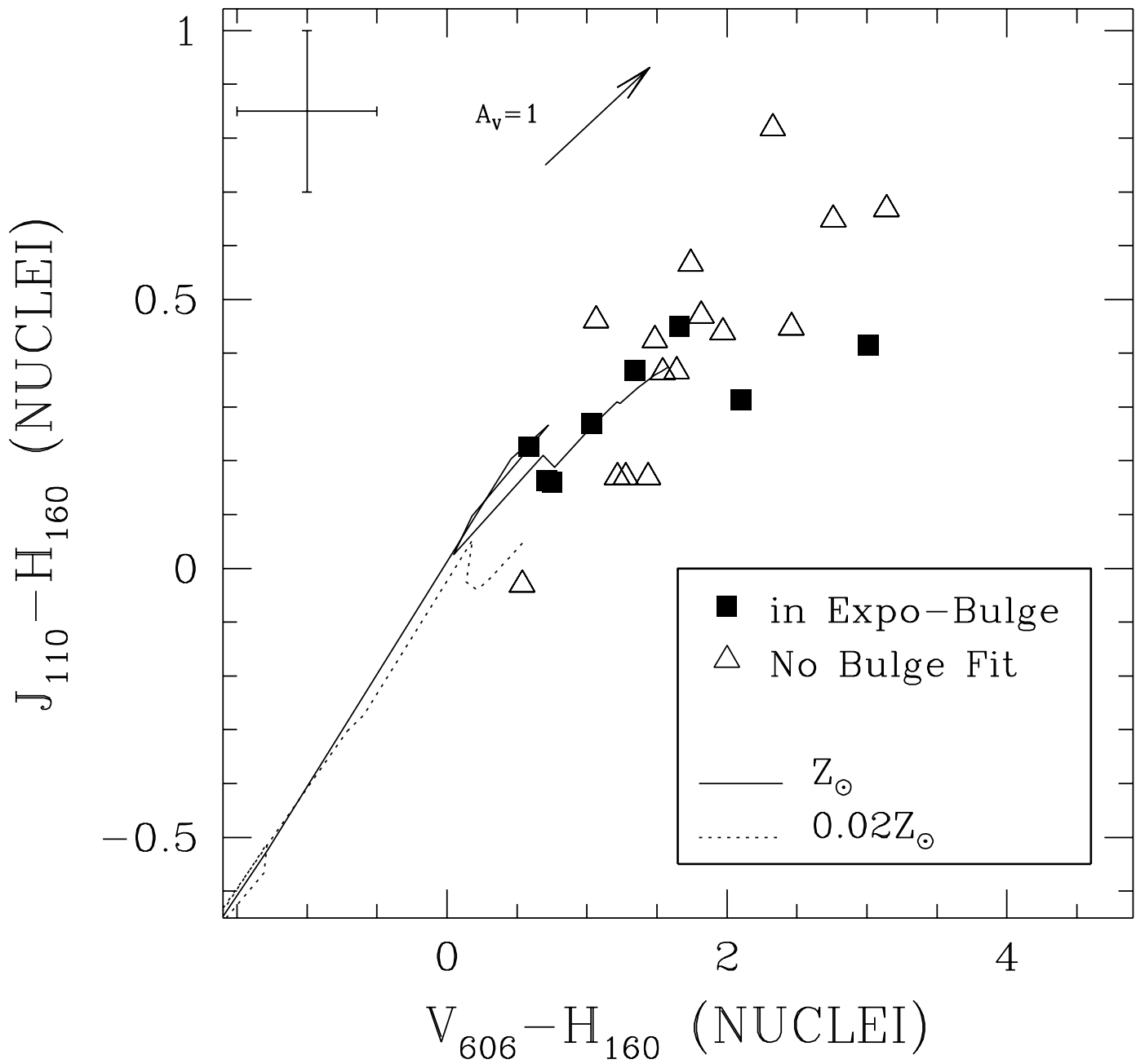
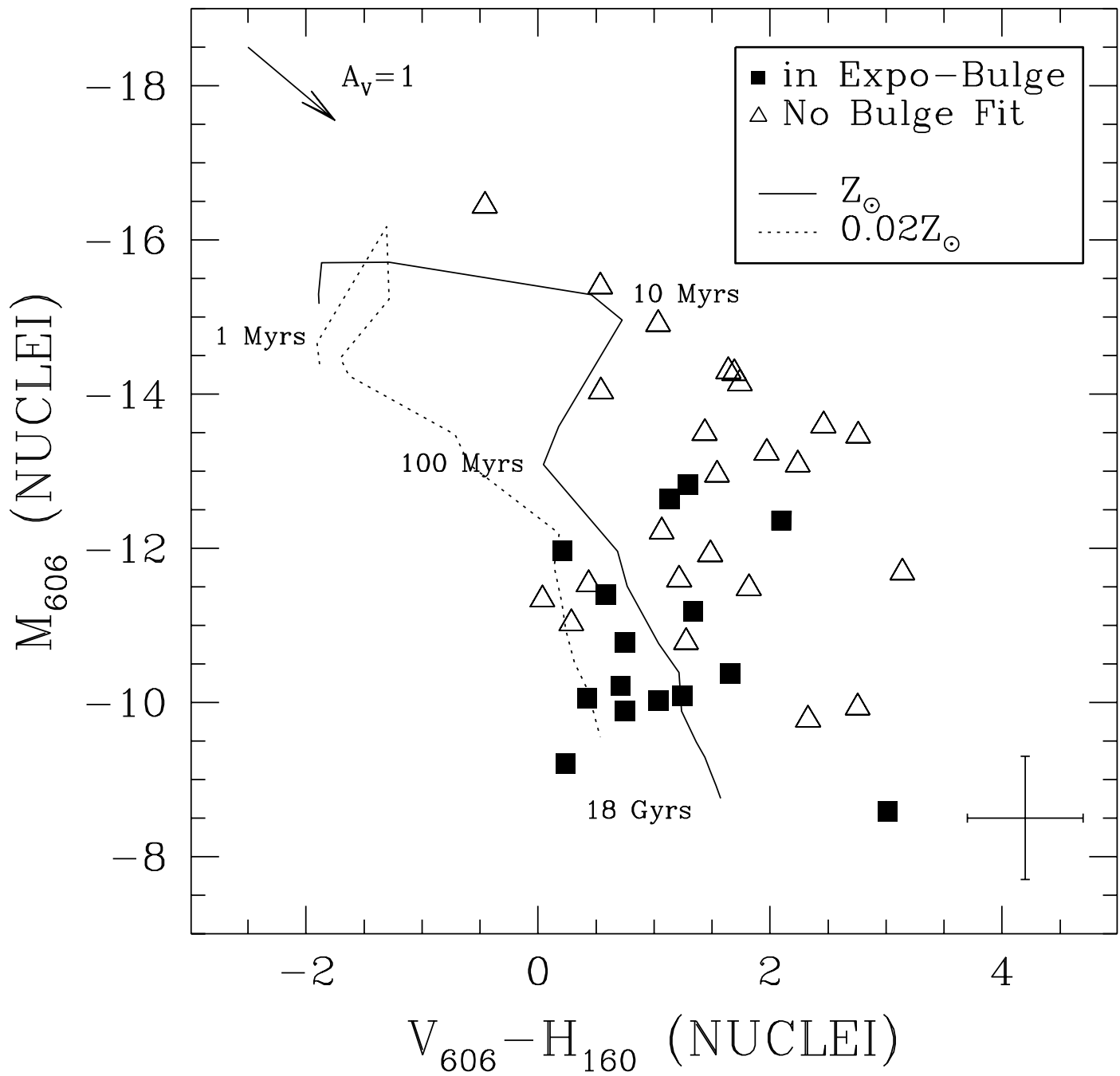
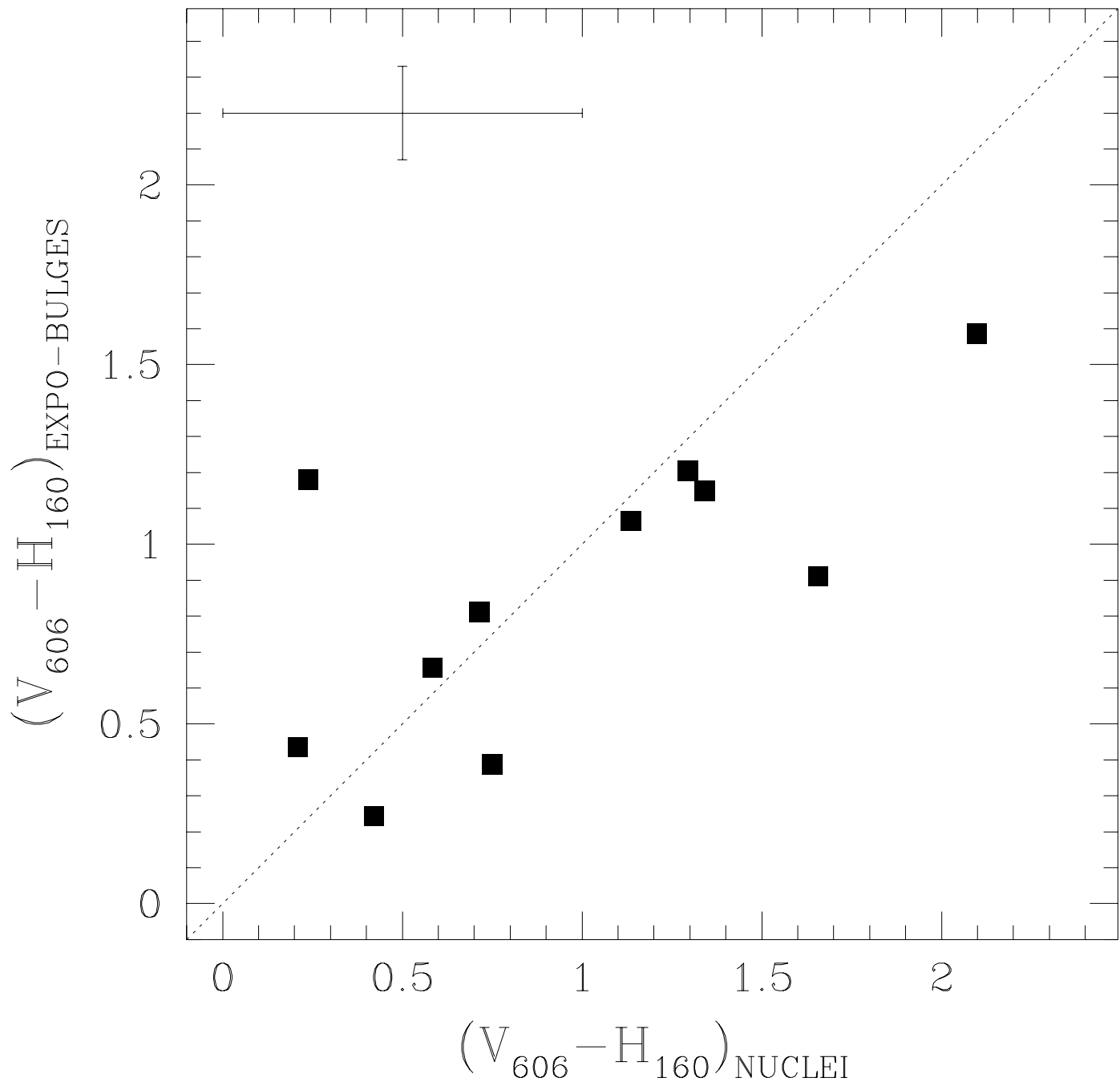
\end{document}